\documentclass[aps,prb,preprint,showpacs,floatfix,superscriptaddress]{revtex4-1}
\usepackage{graphicx,epsfig,dcolumn,amssymb,amsmath}
\usepackage{booktabs}
\usepackage{float}
\usepackage{color}

\begin{document}

\title{Thinning CsPb$_{2}$Br$_{5}$ Perovskite Down to Monolayers: Cs-dependent Stability}

\author{F. Iyikanat}
\email{fadiliyikanat@iyte.edu.tr}
\affiliation{Department of Physics, Izmir Institute of Technology, 35430, Izmir, 
Turkey}

\author{E. Sari}
\affiliation{Department of Photonics, Izmir Institute of Technology, 35430, 
Izmir, Turkey}

\author{H. Sahin}
\email{hasansahin@iyte.edu.tr}
\affiliation{Department of Photonics, Izmir Institute of Technology, 35430, 
Izmir, Turkey}
\affiliation{ICTP-ECAR Eurasian Center for Advanced Research, Izmir Institute of Technology, 35430, Izmir, Turkey}

\date{\today}
\date{\today}

\begin{abstract}

Using first-principles density functional theory calculations, we systematically investigate the structural, electronic and vibrational properties of bulk and potential single-layer structures of perovskite-like CsPb$_{2}$Br$_{5}$ crystal. It is found that while Cs atoms have no effect on the electronic structure, their presence is essential for the formation of stable CsPb$_{2}$Br$_{5}$ crystals. Calculated vibrational spectra of the crystal reveal that not only the bulk form but also the single-layer forms of CsPb$_{2}$Br$_{5}$ are dynamically stable. Predicted single-layer forms can exhibit either semiconducting or metallic character. Moreover, modification of the structural, electronic and magnetic properties of single-layer CsPb$_{2}$Br$_{5}$ upon formation of vacancy defects is investigated. It is found that the formation of Br vacancy (i) has the lowest formation energy, (ii) significantly changes the electronic structure, and (iii) leads to ferromagnetic ground state in the single-layer CsPb$_{2}$Br$_{5}$. However, the formation of Pb and Cs vacancies leads to p-type doping of the single-layer structure. Results reported herein reveal that  single-layer CsPb$_{2}$Br$_{5}$ crystal is a novel stable perovskite with enhanced functionality and a promising candidate for nanodevice applications.
\end{abstract}

\maketitle

\section{Introduction}

In recent years, hybrid organic-inorganic lead halide perovskites have attracted immense interest because of their low cost, easy fabrication and superior optoelectronic 
properties.\cite{Michael, Tan} These hybrid perovskites exhibit great potential in device applications such as photodetectors,\cite{Dou} laser devices,\cite{Zhu} flexible solar cells,\cite{Yang} 
and light emitting diodes (LEDs).\cite{Kim, Xing} One of the most studied organic-inorganic hybrid perovskite is CH$_{3}$NH$_{3}$PbX$_{3}$ (where X = Cl, Br, or I). Noh \textit{et al.} showed that the 
band gap of CH$_{3}$NH$_{3}$Pb(I$_{1-X}$Br$_{X}$) can be controlled by composition engineering and the absorption edge of the mixed halide perovskite can be altered in a controlled manner to cover 
almost the whole visible spectrum.\cite{Noh} It was also demonstrated that organic-inorganic hybrid perovskites exhibit high power conversion efficiency of exceeding 20\%.\cite{Ergen, Jeon} 
The fully inorganic CsPbX$_{3}$ (X = Cl, Br, or I) exhibits higher chemical stability and excellent 
optoelectronic properties compared to organic-inorganic perovskites.\cite{Protesescu, Kulbak, Wang}
It was found that these all-inorganic materials exhibit extremely high fluorescence quantum yield, very narrow emission bandwidth, and suppressed fluorescence blinking.\cite{Swarnkar, Li} It was shown that 
CsPbBr$_{3}$ exhibits high carrier mobility and large diffusion length.\cite{Yettapu} It has been 
theoretically predicted that CsPbBr$_{3}$ is highly defect-tolerant in terms of its electronic 
structure.\cite{Kang}

After successful isolation of graphene monolayers from bulk graphite, research on ultra-thin 2D crystal structures has experienced a remarkable growth.\cite{Novoselov} It has been demonstrated by 
many groups that quantum effects that emerge as a consequence of dimensional reduction may lead to novel features in low-dimensional materials.\cite{Mak, Splendiani} The rapid progress in the synthesis 
and fabrication methods of 2D materials has not only led to exploration of graphene-like materials but also monolayer 2D hybrid perovskites. It was reported that thickness and photoluminescence emission 
of hybrid perovskite nanoplatelets can be controlled by varying the ratio of the organic cations used.\cite{Sichert} Recently, atomically thin organic-inorganic hybrid perovskite synthesized with efficient 
photoluminescence and modulation of color have been achieved by tuning the thickness and composition of the crystal.\cite{Dou1} The synthesis of layered CH$_{3}$NH$_{3}$PbX$_{3}$ (where X = Cl, Br, or I) 
down to a thickness of a few unit cell and even single unit cell layers were accomplished by a combined solution process and vapor-phase conversion method.\cite{Liu} It was found that by controlling 
atomic ratio of the halide anions, stability of the hybrid perovskites can be improved.\cite{Noh} Compared to organic-inorganic hybrid perovskites, all-inorganic perovskites, in which cesium ions 
replace organic cations, exhibit higher chemical stability.\cite{Kulbak,Song} However, low environmental stability of hybrid organic-inorganic perovskites is a crucial issue that needs to be 
addressed for potential future applications. 

Moreover, all-inorganic CsPb$_{2}$Br$_{5}$ emerged as a 2D version of lead halide perovskite materials. CsPb$_{2}$Br$_{5}$ has a tetragonal phase which consists of 
alternating Cs$^{+}$ and [Pb$_{2}$Br$_{5}$]$^{-}$ polyhedron layers. Theoretical and experimental investigations showed that CsPb$_{2}$Br$_{5}$ is an indirect band gap 
semiconductor with a band gap of 2.98 eV.\cite{Li1} Large-scale synthesis of highly luminescent CsPb$_{2}$Br$_{5}$ nanoplatelets was already reported.\cite{Wang1} It 
was found that a dual phase of CsPbBr$_{3}$-CsPb$_{2}$Br$_{5}$ exhibits increased conductivity and improved emission life time compare to that of the pure 
CsPbBr$_{3}$.\cite{Zhang} However, thickness and composition dependent structural stability and electronic 
properties of CsPb$_{2}$Br$_{5}$ remain almost unexplored. It is known that ultrathin 2D materials are more 
sensitive to environmental conditions than their bulk counterpart. Environmental conditions and substrate can 
affect the structural stability and characteristic properties of ultrathin 2D materials. It was reported that 
in the presence of chlorine, when CH$_{3}$NH$_{3}$PbI$_{3}$ films are deposited on a TiO$_{2}$ mesoporous 
layer, a stable cubic phase is formed in CH$_{3}$NH$_{3}$PbI$_{3}$ perovskite, instead of the commonly 
observed tetragonal phase.\cite{Wang2} In a recent study CsPb$_{2}$Br$_{5}$ nanosheets were obtained via an 
oriented attachment of CsPbBr$_{3}$ nanocubes.\cite{Li1} It was also reported that inorganic Cs atom plays a 
significant role in the stability of perovskites.\cite{Zhang1, Saliba} The Cs atom as a cation, donates charge 
to the lattice of a perovskite and fulfill the charge neutrality.

Although the bulk forms of many perovskites have been studied intensely, effect of dimensional reduction in the characteristic properties of CsPb$_{2}$Br$_{5}$ 
crystal has not been investigated before. In the current study, using first principles calculations based on density functional theory (DFT), we present a detailed analysis 
of the structural, electronic, vibrational and vacancy-dependent characteristics of single-layer CsPb$_{2}$Br$_{5}$. The paper is organized as follows. Details of the 
computational methodology are given in Sec. \ref{sec2}. Structural and electronic properties of bulk CsPb$_{2}$Br$_{5}$ are given in Sec. \ref{sec3}. The effect of 
dimensional crossover on the structure and electronic properties of CsPb$_{2}$Br$_{5}$ crystal is presented in Sec. \ref{sec4}. Structural, electronic, and magnetic 
properties of various vacancy defects are discussed in Sec. \ref{sec5}. We conclude our results in Sec. \ref{sec6}.

\begin{figure}
\includegraphics[width=12 cm]{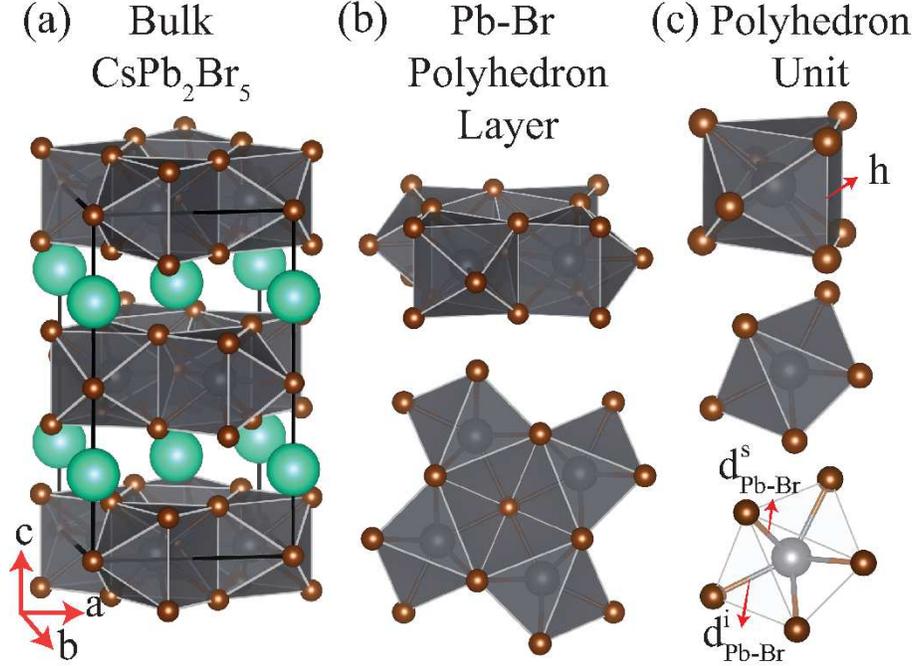}
\caption{\label{fig-1}
(Color online) 
(a) Tilted side view of bulk CsPb$_{2}$Br$_{5}$. (b) Tilted side and top views of Pb-Br polyhedron layer. (c) Tilted side and top views of Pb-Br polyhedron. Green, brown, and gray balls illustrate Cs, Br, and Pb atoms, respectively.}
\end{figure}

\section{Computational Methodology}\label{sec2}

All the calculations were performed using the projector augmented wave (PAW)\cite{Paw1, Paw2} potentials as implemented in the Vienna \textit{ab initio} Simulation Package (VASP).\cite{Vasp1, Vasp2} The local 
density approximation (LDA)\cite{Perdew} was used with the inclusion of spin-orbit coupling (SOC) to describe exchange and correlation potential as parametrized by the Ceperley and Alder functional.\cite{Ceperley} 
To obtain partial charge on the atoms, the Bader technique was used.\cite{Henkelman}

A plane-wave basis set with kinetic energy cutoff of 500 eV was used for all the calculations. The total energy difference between the sequential steps in the iterations was taken as 10$^{-5}$ eV as a convergence 
criterion. The total force in the unit cell was reduced to a value of less than 10$^{-4}$ eV/\AA{}. For bulk and single-layer CsPb$_{2}$Br$_{5}$ $\Gamma$-centered \textit{k}-point meshes of 4$\times$4$\times$3 and 
4$\times$4$\times$1 were used, respectively. In order to model single-layer CsPb$_{2}$Br$_{5}$, a vacuum spacing larger than 14 \AA{} was inserted to avoid spurious interactions between adjacent layers. 
Dipole corrections were applied in the direction perpendicular to the plane of the 
charged monolayers.\cite{Makov} Gaussian smearing of 0.05 eV was used for electronic density of states 
calculations. The cohesive energy per atom was formulated as

\begin{equation}
 E_{coh}=\left[\sum n_{a}E_{a}-E_{Str}\right]/N
\end{equation}
where \textit{E$_{a}$} denotes the energy of a single isolated atom and \textit{n$_{a}$} denotes the number of corresponding atoms contained in the unit cell. \textit{E$_{Str}$} denotes the total energy of the 
structure. \textit{N} denotes the number of total atoms contained in the unit cell. Phonon calculations were performed by making use of the small displacement method as implemented in the PHON software 
package.\cite{Alfe} Out-of plane acoustic phonon modes in CsPb$_{4}$Br$_{10}$, 
\textit{\textbf{ss}}-CsPb$_{2}$Br$_{5}$ and \textit{\textbf{ds}}-CsPb$_{2}$Br$_{5}$ crystal structures are 
corrected by quadratic-fitting at the vicinity of zone center. (See also SI Fig. S2 (a) and (b).)

\begin{figure}
\includegraphics[width=11 cm]{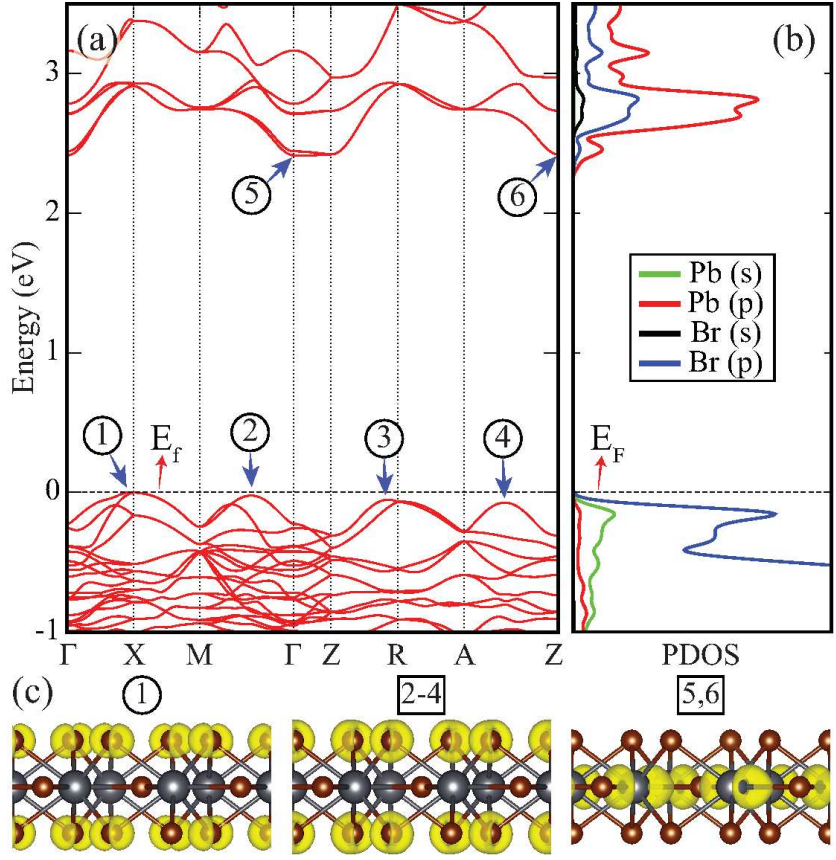}
\caption{\label{fig-2}
(Color online) 
(a) The energy-band dispersion and (b) partial density of states of bulk 
CsPb$_{2}$Br$_{5}$. (c) The band decomposed charge densities 
of the bulk CsPb$_{2}$Br$_{5}$ at the labeled band edges. Isosurface value of charge density is 
2$\times$10$^{-5}$ \textit{e}/\AA{}$^{3}$.}
\end{figure}

\section{Bulk 
C\MakeLowercase{s}P\MakeLowercase{b}$_{2}$B\MakeLowercase{r}$_{5}$}\label{sec3}

\begin{table*}
\caption{\label{table1}The calculated ground state properties for bulk (\textit{\textbf{b}}) 
CsPb$_{2}$Br$_{5}$ and single-layer Pb$_{2}$Br$_{5}$, CsPb$_{4}$Br$_{10}$, \textit{\textbf{ss}}-CsPb$_{2}$Br$_{5}$, 
\textit{\textbf{ds}}-CsPb$_{2}$Br$_{5}$, and Cs$_{2}$Pb$_{2}$Br$_{5}$: The 
lattice constants in the lateral and vertical directions, \textit{a} and \textit{c}, respectively; thickness of the Pb-Br 
polyhedron layer, \textit{h}; atomic distances between Pb and Br atoms, 
d$^{i}_{Pb-Br}$ and d$^{s}_{Pb-Br}$; atomic distance between Cs and 
surface Br atoms, d$^{s}_{Cs-Br}$; the cohesive energies, E$_{coh}$; the amount of donated ($-$) charge by the Br$_{i}$, Br$_{s}$, Pb, and Cs atoms 
are $\Delta\rho^{i}_{Br}$, $\Delta\rho^{s}_{Br}$, $\Delta\rho_{Pb}$, and $\Delta\rho_{Cs}$, respectively; the energy band gap of the 
structure, Gap; the work function, $\Phi$.}
\begin{tabular}{lcccccccccccccc}
\hline\hline
                            &     \textit{a}   &    \textit{c}    & \textit{h}                 &  d$^{i}_{Pb-Br}$     &     d$^{s}_{Pb-Br}$    &     d$^{s}_{Cs-Br}$  &  E$_{coh}$ & $\Delta\rho^{i}_{Br}$ &  
$\Delta\rho^{s}_{Br}$ &  $\Delta\rho_{Pb}$ & $\Delta\rho_{Cs}$ & Gap      & $\Phi$  
 \\
                            & (\AA{}) & (\AA{}) &(\AA{})            &  (\AA{})   
        & (\AA{})           &     (\AA{})   
      &    (eV)    & (\textit{e})       & (\textit{e})        & 
(\textit{e})       & (\textit{e}) & (eV)  &    (eV)  \\
\hline
\textit{\textbf{b}} CsPb$_{2}$Br$_{5}$        &   8.24  & 14.57  & 3.83                &    3.08   
         &      2.92           & 3.53                 
       & 3.42       &    0.6             &   0.6               &   -1.0  
           &    -0.8           & 2.41          & 1.17         \\ 
Pb$_{2}$Br$_{5}$         &   8.09  &    -   &   3.72              &   3.02       
      &   2.88           &      -         
      &  3.08      &    0.5             &   0.4               &   -1.0    
         &    -              &    -               & 7.00         \\ 
CsPb$_{4}$Br$_{10}$         &   8.18  &    -   &   3.77              &   3.05, 3.08 
             &   2.94         &      3.44        &  
  3.21      &    0.5             &   0.5               &   -1.0    
         &    -0.8              &      -        & 4.53        \\ 
\textit{\textbf{ss}}-CsPb$_{2}$Br$_{5}$       &   8.29  &    -   &    3.85             &   3.11       
       &   2.84           & 3.42          & 
  3.31      &    0.6             &   0.5               &   -1.0    
         &     -0.8          & 2.53          & 2.75         \\ 
\textit{\textbf{ds}}-CsPb$_{2}$Br$_{5}$       &   8.29  &    -   &    3.81             &   3.11       
         &   2.92           & 3.43          & 
  3.32      &    0.6             &   0.6               &   -1.0    
         &     -0.8          & 2.54          & 4.79         \\ 
Cs$_{2}$Pb$_{2}$Br$_{5}$ &   8.21  &    -   &   4.31              &   3.18       
     &   3.22           & 3.35          & 
  3.15      &    0.6             &   0.7               &   -0.8    
         &   -0.8            & -             & 1.45         \\ 
\hline\hline 
\end{tabular}
\end{table*}

Before a detailed discussion of the potential single-layer forms of CsPb$_{2}$Br$_{5}$, we first investigate structural properties of bulk form of the crystal. The tetragonal phase of CsPb$_{2}$Br$_{5}$ is shown 
in Fig \ref{fig-1} (a). CsPb$_{2}$Br$_{5}$ consists of sandwiched structure with alternating Cs and Pb-Br polyhedron layers. The calculated lattice parameters are \textit{a} $=$ 8.24 \AA{} and \textit{c} 
$=$ 14.57 \AA{}. Tilted side and top views of Pb-Br polyhedron layer are shown in Fig \ref{fig-1} (b). Pb-Br polyhedron layer is composed of Pb-Br polyhedrons which are formed by putting a triangular prism 
and two rectangular pyramids together. Tilted side and top views of Pb-Br polyhedron are shown Fig \ref{fig-1} (c). There are two different bromine atoms in the polyhedron: one of them is located at the 
surface of Pb-Br polyhedron layer (Br$_{s}$) and the other one is located at the inner plane of the layer (Br$_{i}$). Br$_{i}$ is located in the same plane with Pb atoms and it has the Pb-Br bond distance 
of d$^{i}_{Pb-Br}$ $=$ 3.08 \AA{}. The bond length between Br$_{s}$ and Pb atoms is d$^{s}_{Pb-Br}$ $=$ 2.92 \AA{}. The thickness of Pb-Br polyhedron layer is \textit{h} $=$ 3.83 \AA{}. The distance between 
Cs and surface Br atoms is d$^{s}_{Cs-Br}$ $=$ 3.53 \AA{}. It is found that the cohesive energy per atom of CsPb$_{2}$Br$_{5}$ crystal is 3.42 eV.

For detailed investigation of the bondings in CsPb$_{2}$Br$_{5}$ crystal Bader charge analysis is also performed. As given in the Table \ref{table1}, 0.8 and 1.0 \textit{e} charges are donated by each Cs and 
Pb atoms, respectively. On the other hand, each Br atom receives 0.6 \textit{e} charge. Therefore, analysis of the electronic structure reveals that ionic interaction arises between the layers of Cs and Pb-Br 
skeleton via vertical charge transport from Cs layer to the Pb-Br skeleton. Furthermore, the bond between Pb and Br atoms in the Pb-Br skeleton has also an ionic character. Cs terminated surface of CsPb$_{2}$Br$_{5}$ 
crystal has a low work function of $\Phi$ $=$ 1.17 eV.

Fig. \ref{fig-2} shows the band structure and projected density of states (PDOS) for CsPb$_{2}$Br$_{5}$ crystal. It can clearly be seen from the figure, that bulk CsPb$_{2}$Br$_{5}$ exhibits an indirect band gap, 
with the valence band maximum (VBM) residing along the line $\Gamma$-X ($\textcircled{1}$), while the conduction band minimum (CBM) being located at the $\Gamma$ point ($\textcircled{5}$). The indirect gap 
calculated with LDA+SOC is 2.41 eV. It is known that experimentally observed bandgaps, which are 
underestimated by both bare-LDA and bare-GGA functionals, can be well-approximated by the considering 
screening (GW calculations) and excitonic (solving Bethe-Salpeter equation) effects. However, due to high 
computational cost, GW and BSE calculations are not taken into account here. The partial contributions of 
orbital states to electronic DOS of bulk CsPb$_{2}$Br$_{5}$ crystal are given in the Fig. \ref{fig-2} (b). It 
appears that the major contribution to the states around the band edges originates from Pb and Br atoms. While 
the VBM is dominated by \textit{p} orbitals of Br atom, the CBM is mostly composed of the $p$ orbitals of Pb 
atom. The states of the Cs atom reside at deep energy levels, therefore, they have no effect on the 
electronic structure of the crystal. Band-decomposed charge densities of the valence and conduction band 
edges are also given in the Fig. \ref{fig-2}. As seen in the figure, edges in the top of valence band are 
composed of the \textit{p$_{x}$} and \textit{$p_{y}$} orbitals of Br$_{s}$ atoms. Valance band edges in 
between $\Gamma$ and M ($\textcircled{2}$), Z and R ($\textcircled{3}$), A and Z ($\textcircled{4}$) points 
differ by 23, 
53 and 74 meV energy than the VBM of the single-layer, respectively. On the other hand, top of conduction band edges ($\textcircled{5}$ and $\textcircled{6}$) are mostly made up of \textit{$p_{x}$} and 
\textit{$p_{y}$} orbitals of Pb atom.

\begin{figure*}
\includegraphics[width=17 cm]{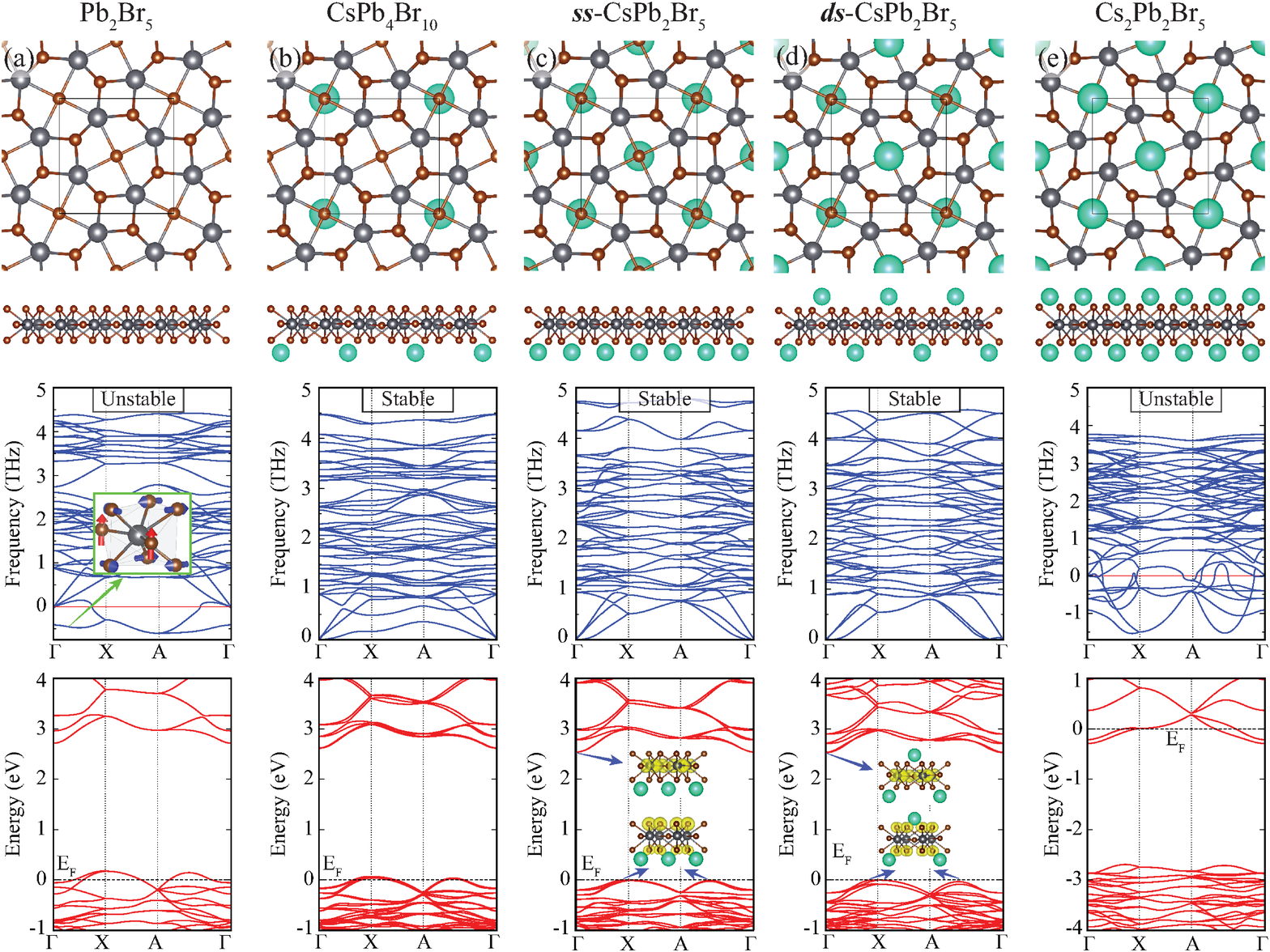}
\caption{\label{fig-3}
(Color online) 
Possible single-layer structures of CsPb$_{2}$Br$_{5}$ crystal. Top and side 
views of the structures, phonon spectrum, and SOC included electronic band diagrams of single-layer 
(a) Pb$_{2}$Br$_{5}$, (Inset: Tilted side view of atomic displacements of the corresponding mode.) (b) CsPb$_{4}$Br$_{10}$,
(c) \textit{\textbf{ss}}-CsPb$_{2}$Br$_{5}$, (d) \textit{\textbf{ds}}-CsPb$_{2}$Br$_{5}$, and (e) Cs$_{2}$Pb$_{2}$Br$_{5}$. Isosurface value 
of band decomposed charge densities (inset of band structure of \textit{\textbf{ss}}-CsPb$_{2}$Br$_{5}$) 
is 2$\times$10$^{-5}$ \textit{e}/\AA{}$^{3}$.}
\end{figure*}

\section{Thinning C\MakeLowercase{s}P\MakeLowercase{b}$_{2}$B\MakeLowercase{r}$_{5}$ down to 
Monolayers}\label{sec4}

In this section, we investigate possible stable structure of single-layer CsPb$_{2}$Br$_{5}$ crystal and the effect of dimensional crossover on its characteristic properties. Five reasonable single-layer 
configurations derived from bulk  CsPb$_{2}$Br$_{5}$ structure are shown in Fig. \ref{fig-3}. According to their chemical compositions, these five structures are named as Pb$_{2}$Br$_{5}$, CsPb$_{4}$Br$_{10}$, 
single-side Cs terminated CsPb$_{2}$Br$_{5}$ (\textit{\textbf{ss}}-CsPb$_{2}$Br$_{5}$), double-side Cs terminated CsPb$_{2}$Br$_{5}$ (\textit{\textbf{ds}}-CsPb$_{2}$Br$_{5}$) and Cs$_{2}$Pb$_{2}$Br$_{5}$. Depending on whether the synthesis 
technique is growth or mechanical exfoliation process one of these of the single layer crystal structures or their mixtures can be obtained.

Single-layer Pb$_{2}$Br$_{5}$ is a Cs-free skeleton composed of only Pb-Br polyhedrons. As given in Table \ref{table1}, the calculated lattice parameter and thickness of Pb$_{2}$Br$_{5}$ are \textit{a} $=$ 8.09 \AA{} 
and \textit{h} $=$  3.72 \AA{}, respectively. These values are relatively small compared to those of bulk. The calculated bond lengths between the atoms of the structure are given in the Table \ref{table1}. To assess 
the dynamical stability, phonon dispersions of single-layer Pb$_{2}$Br$_{5}$ are calculated and shown in Fig. \ref{fig-3} (a). A 3$\times$3$\times$1 supercell is used for the phonon band structure calculations of 
single-layer forms of the crystal. The phonon spectra of the structure exhibit imaginary eigenvalues through all the symmetry points, indicating the instability of the structure. Instability in the structure could be attributed 
to the charge transfer mechanism between ions of the structure. As given in Table \ref{table1}, resulting charges of Br atoms of Pb$_{2}$Br$_{5}$ are lower than that of bulk CsPb$_{2}$Br$_{5}$. Although each Pb atom 
donates 1.0 \textit{e} charge, each Br$_{i}$ and Br$_{s}$ atoms receive 0.5 and 0.4 \textit{e} charges, respectively. Therefore, unsaturated Br$_{i}$ and Br$_{s}$ atoms may lead to dynamical instability in 
single-layer Pb$_{2}$Br$_{5}$.

For a deeper analysis of instability of single-layer Pb$_{2}$Br$_{5}$, atomic displacements of the lowest mode are shown in the inset of phonon band diagram of Fig. \ref{fig-3} (a). It appears that the 
mode responsible for the instability of the structure is due to vibrations of Br$_{i}$ and Br$_{s}$ atoms normal and parallel to the plane of the structure, respectively. Imaginary eigenfrequencies in the whole Brillouin 
Zone indicate the lack of required restoring force against these atomic motions. Therefore, the structure is unstable under this motion and transforms to another phase. This instability can be cured by increasing the 
bond strength between Pb and Br atoms through adding an extra charges.

As shown in the Fig. \ref{fig-3} (b), single-layer of CsPb$_{4}$Br$_{10}$ is formed through half of the single-side of Pb-Br layer covered by Cs atoms. The lattice parameter 
and the thickness of the structure are \textit{a} $=$ 8.18 \AA{} and \textit{h} $=$  3.77 \AA{}, respectively. It is found that the single-layer CsPb$_{4}$Br$_{10}$ is 
formed by a cohesive energy of 3.21 eV.  Since one side of the structure is covered by Cs atoms, the bonds show anisotropy in the vertical direction (see Table \ref{table1}). 
Given values of the bond lengths belong to Cs terminated side of the structure. The phonon spectra of single-layer CsPb$_{4}$Br$_{10}$ exhibit real eigenvalues through 
all the symmetry points, confirming the dynamical stability of the layers. Bader charge analysis shows that Cs and Pb atoms of the structure donate 0.8 and 1.0 \textit{e} charges, respectively. 
Therefore the amount of received charges by each Br atom is 0.5 \textit{e}. The work function of the structure is calculated to be 4.53 eV. It is also seen from the 
Fig. \ref{fig-3} (b) that lowering the concentration of Cs atoms leads to unfilling of the valence band edge and therefore p-type conductivity in the single-layer 
CsPb$_{4}$Br$_{10}$.

Relaxed geometric structures of \textit{\textbf{ss}}-CsPb$_{2}$Br$_{5}$ and \textit{\textbf{ds}}-CsPb$_{2}$Br$_{5}$ are shown in Figs. \ref{fig-3} (c) and (d), respectively. These two structures possess the same 
chemical composition with the CsPb$_{2}$Br$_{5}$ crystal. The lattice parameters of both of the structures are \textit{a} $=$ 8.29 \AA{}. The thickness of the Pb$_{2}$Br$_{5}$ sublayer of \textit{\textbf{ss}}-CsPb$_{2}$Br$_{5}$ 
and \textit{\textbf{ds}}-CsPb$_{2}$Br$_{5}$ are \textit{h} $=$ 3.85 and 3.81 \AA{}, respectively. As given in the Table \ref{table1}, the atom-atom bond distances of these two structures are very similar 
to that of the bulk CsPb$_{2}$Br$_{5}$. Real frequencies in the phonon spectra of single-layer structures (Figs. \ref{fig-3} (c) and (d)) of \textit{\textbf{ss}}-CsPb$_{2}$Br$_{5}$ and 
\textit{\textbf{ds}}-CsPb$_{2}$Br$_{5}$ appear in a whole Brillouin Zone, indicating the dynamically stabilities of these two structures. Bader charge analysis reveals that the charge transfer mechanism 
of single-layers of \textit{\textbf{ss}}-CsPb$_{2}$Br$_{5}$ and \textit{\textbf{ds}}-CsPb$_{2}$Br$_{5}$ are very similar to the bulk CsPb$_{2}$Br$_{5}$. Cs and Pb atoms of the both structures donate 0.8 
and 1.0 \textit{e} charges, respectively. The amount of charge received by Br atoms of the two structures are in the range of 0.5-0.6 \textit{e}. Therefore, Pb-Br skeletons of the two structures are saturated 
with enough charges to maintain stable bonding mechanism. In addition, cohesive energies of these two structures are given in Table \ref{table1}. Among the single-layer forms of the crystal, due to the vertical 
symmetry and the same chemical composition with the bulk counterpart, \textit{\textbf{ds}}-CsPb$_{2}$Br$_{5}$ has the highest cohesive energy. Furthermore, calculated work function values of singe-layers 
of \textit{\textbf{ss}}-CsPb$_{2}$Br$_{5}$ and \textit{\textbf{ds}}-CsPb$_{2}$Br$_{5}$ are 2.75 and 4.79 eV, respectively. Calculated values of the work functions belong to Cs-terminated side of the structure. 
Hence, the work function shows a decreasing behavior with increasing concentration of Cs atoms of Cs-terminated side. As shown in Fig. \ref{fig-3} (c) and (d), single-layer \textit{\textbf{ss}}-CsPb$_{2}$Br$_{5}$ 
and \textit{\textbf{ds}}-CsPb$_{2}$Br$_{5}$ have very similar electronic band diagrams and they are indirect band gap semiconductors. The VBMs of the two structures reside at X point, whereas their CBMs reside at 
$\Gamma$ point. It is worth mentioning that valence band edge in between A and $\Gamma$ points differs only by 24 meV energy from the VBM of the two structure. Band  decomposed charge densities at the valence band and the conduction band edges of the two structures are shown in Figs. \ref{fig-3} (c) and (d). Band edge characteristics of these structures possess very similar behavior with their bulk counterpart. As given in Table \ref{table1}, the SOC included band gap of \textit{\textbf{ss}}-CsPb$_{2}$Br$_{5}$ and \textit{\textbf{ds}}-CsPb$_{2}$Br$_{5}$ are 2.53 and 2.54 eV therefore with decreasing thickness from bulk to single-layer, the band gap of CsPb$_{2}$Br$_{5}$ increases by $\sim$ 0.13 eV.

The single-layer Cs$_{2}$Pb$_{2}$Br$_{5}$ is constructed by three submonolayers: a Pb$_{2}$Br$_{5}$ sublayer and two surrounding sublayers of Cs atoms. The calculated lattice parameter of Cs$_{2}$Pb$_{2}$Br$_{5}$, \textit{a} $=$ 8.21 \AA{} is very similar to bulk lattice parameter of the crystal. However, the thickness of Pb$_{2}$Br$_{5}$ sublayer is \textit{h} $=$ 4.31 \AA{}, which is much higher than that of the other single-layer forms of the crystal.  As in Pb$_{2}$Br$_{5}$, phonon spectra of single-layer Cs$_{2}$Pb$_{2}$Br$_{5}$, shown in Fig. \ref{fig-3} (e), exhibit imaginary eigenvalues through whole the symmetry points, indicating the dynamical instability of the structure. In single-layer Cs$_{2}$Pb$_{2}$Br$_{5}$, while each Cs and Pb atom donate 0.8 \textit{e} charges, each Br$_{i}$ and Br$_{s}$ atoms receive 0.6 and 0.7 \textit{e} charges, respectively. Extra charges provided by Cs atoms lead to a decrease in charge transfer between Pb and Br atoms. Therefore, the Cs-induced weakening of the Pb-Br bonds is the reason of the instability in the single-layer Cs$_{2}$Pb$_{2}$Br$_{5}$.

\subsection{Effect of C\MakeLowercase{s} Atoms via The Charging of P\MakeLowercase{b}$_{2}$B\MakeLowercase{r}$_{5}$ Skeleton}

In this chapter, in order to facilitate a deeper understanding of the role of Cs atoms, we examine the stability of Cs-free Pb$_{2}$Br$_{5}$, which is already presented to be unstable in the previous chapter, via charging calculations.

One low-lying optical and one acoustic phonon branches that have imaginary eigenfrequencies are the direct indication of dynamical instability in the single layer structure of Cs-free Pb$_{2}$Br$_{5}$. However, as calculated in the previous chapter, single layer skeleton structure of Pb$_{2}$Br$_{5}$ can be stabilized through the adsorption of Cs atoms.  Therefore, the question arises as to whether or not the stability of CsPb$_{2}$Br$_{5}$ is provided only by the charge transferred from Cs atoms to Pb$_{2}$Br$_{5}$ skeleton.  

\begin{figure}
	\includegraphics[width=15 cm]{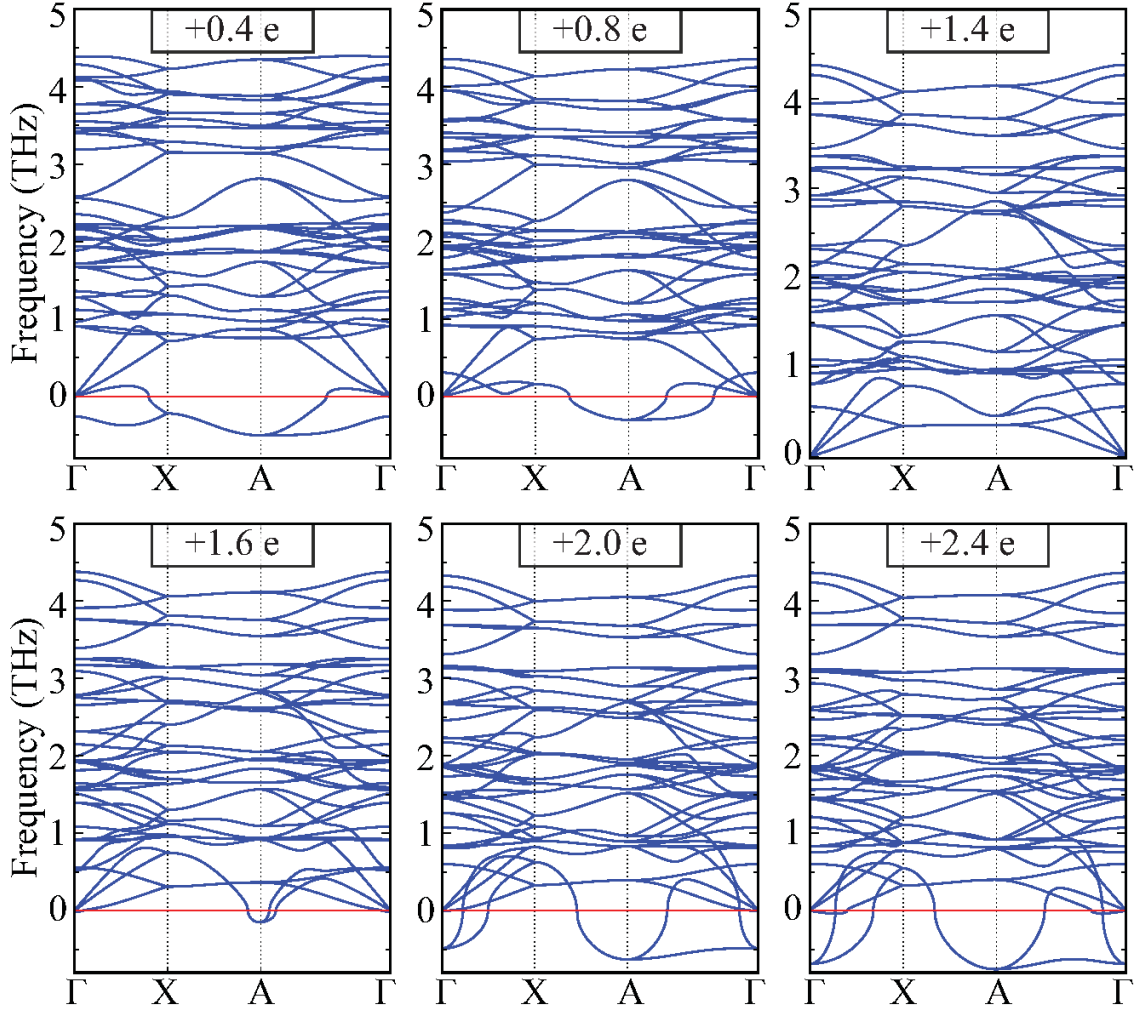}
	\caption{\label{fig-45}
		(Color online) Charging-dependent phonon dispersion of single layer skeleton structure of Pb$_{2}$Br$_{5}$}
\end{figure}

As shown in Fig. \ref{fig-45}, the role of electron transfer on the stability of Pb$_{2}$Br$_{5}$ can be examined through the addition of extra electrons into the primitive unit cell. Here, it is important to note that once the unit cell is charged by extra electrons, structure is re-optimized by considering the new charge distribution and phonon calculations are performed for this fully-relaxed structure.  Fig. \ref{fig-45} shows that one acoustic and one low-lying optical branches, having imaginary eigenfrequencies, of the Pb$_{2}$Br$_{5}$ skeleton can be fixed upon the charging. It appears that while the structure is fully stabilized under 1.4 e charging, it exhibits instability under over-charged and less-charged situations. Therefore, it can be concluded that Cs atoms play an important role as stabilizer of the Pb$_{2}$Br$_{5}$ skeleton by charge transfer. Such a charge dependent stability may pave the way to synthesize novel Cs-free perovskite structures and to fix the vacancy-dependent instabilities in similar structures.   

\begin{figure*}
	\includegraphics[width=17 cm]{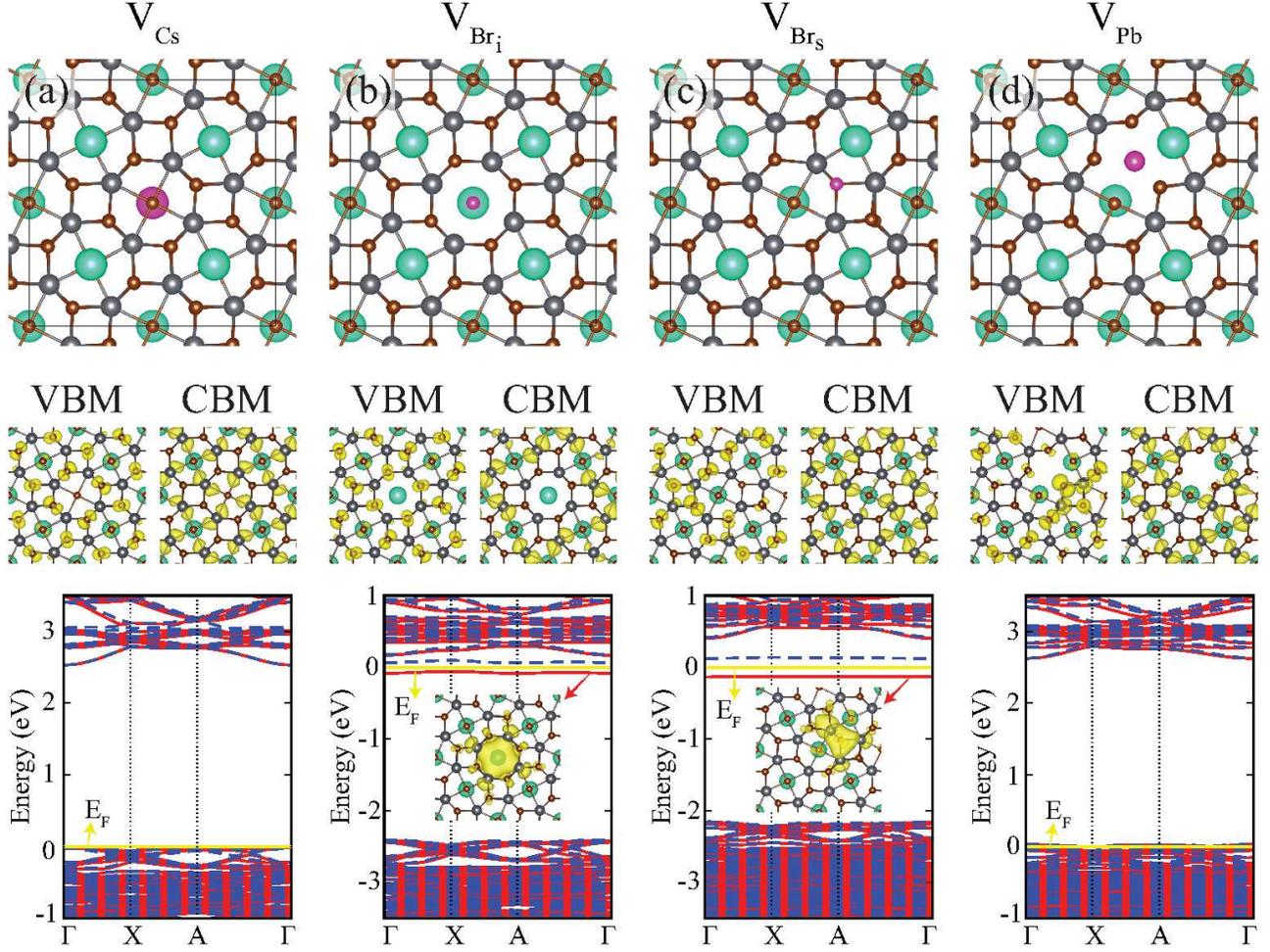}
	\caption{\label{fig-4}
		(Color online) 
		Top view of the structures, band decomposed charge densities of the VBM and CBM, 
		and electronic band structures of single-layer 
		\textit{\textbf{ds}}-CsPb$_{2}$Br$_{5}$ with (a) Cs vacancy, (b) Br$_{i}$ vacancy, (c) Br$_{s}$ 
		vacancy, and (d) Pb vacancy. Pink atoms illustrate removed atoms. Yellow, red, 
		and dashed blue lines in band diagrams illustrate the Fermi level, majority, 
		and minority spin bands, respectively. Band decomposed charge densities of
		in-gap states of V$_{Br_{i}}$ and V$_{Br_{s}}$ are shown in the insets of the corresponding band diagrams.
		Isosurface value of charge density is 6$\times$10$^{-6}$ \textit{e}/\AA{}$^{3}$.}
\end{figure*}

\section{Vacancy Defects in Single-Layer 
C\MakeLowercase{s}P\MakeLowercase{b}$_{2}$B\MakeLowercase{r}$_{5}$}\label{sec5}

During the growth or exfoliation of layered crystals, existence of various lattice 
imperfections is inevitable. One of the most common lattice imperfection in 
layered crystals is vacancies. The formation of vacancies in ionically bonded materials are the source of trap states and profoundly alter the electronic and optical properties of semiconductors. Therefore, it is important to 
investigate how vacancies are formed and what are the characteristic features. 
In this section, four different vacancy types in single-layer 
\textit{\textbf{ds}}-CsPb$_{2}$Br$_{5}$ that corresponds to the energetically the most favorable structure
are considered, Cs vacancy (V$_{Cs}$), Br$_{i}$ vacancy (V$_{Br_{i}}$), Br$_{s}$ vacancy (V$_{Br_{s}}$), and Pb vacancy (V$_{Pb}$). 
In order to hinder the interaction between vacancies in adjacent cells, a 
64-atom supercell was used. Relative stabilities of the four 
vacancies are calculated according to the formula: 
\begin{equation}
E_{form} = E_{SL+vac}+E_{A}-E_{SL}
\end{equation}
where \textit{E$_{form}$} is the formation energy of the relevant vacancy, 
\textit{E$_{SL+vac}$} is the total energy of the supercell with vacancy, 
\textit{E$_{A}$} is the isolated-single-atom energy of the removed atom and \textit{E$_{SL}$} is the 
total energy of supercell of single layer.

\begin{table}[htbp]
\caption{\label{table3} The lattice constants, \textit{a}; formation energies, 
E$_{form}$; magnetic moments, M; electronic characteristics, and band gaps of 
four different defected forms of 2$\times$2$\times$1 supercell of the single-layer 
CsPb$_{2}$Br$_{5}$.}
\begin{tabular}{lccccc}
\hline\hline
                         & a          & E$_{form}$ &     M      &    Electronic  
  & Band Gap   \\
                         &  (\AA{})     &   (eV)     & ($\mu_{B}$) &  
Characteristic  &   (eV)     \\
\hline
V$_{Cs}$                 &  16.48     &   4.82     &     0.0     &   
semiconductor &    2.53    \\ 
V$_{Br_{i}}$               &   16.44     &   5.02    &     1.0     &    
semiconductor &    0.12    \\ 
V$_{Br_{s}}$               &   16.48     &   4.51    &     1.0     &    
semiconductor &    0.25    \\ 
V$_{Pb}$                &   16.46     &   8.20    &     0.0     &       metal    
  &     -       \\ 
\hline\hline 
\end{tabular}
\end{table}

Relaxed geometric structures when a single Cs, Br$_{i}$, Br$_{s}$ and Pb are created in the \textit{\textbf{ds}}-CsPb$_{2}$Br$_{5}$ are shown in Figs. 
\ref{fig-4} (a)-(d), respectively. Formation of these vacancies leads to negligible distortion on the single-layer \textit{\textbf{ds}}-CsPb$_{2}$Br$_{5}$. 
As given in the Table \ref{table1} fully relaxed lattice parameters of V$_{Cs}$, V$_{Br_{i}}$, V$_{Br_{s}}$ and V$_{Pb}$ are 16.48, 16.44, 16.48 and 16.46 \AA{}, respectively. The formation of Pb vacancy (V$_{Pb}$) leads to minor local reconstructions. When Pb vacancy is introduced, the closest Br atoms to the extracted Pb atom are released and they move toward the neighboring Pb atoms. The 
formation energies of V$_{Cs}$, V$_{Br_{i}}$, V$_{Br_{s}}$ and V$_{Pb}$ are 
calculated to be 4.82, 5.02, 4.51 and 8.20 eV, respectively. Therefore, one can conclude 
that formation of Cs and Br vacancies in single-layer \textit{\textbf{ds}}-CsPb$_{2}$Br$_{5}$ perovskite are more likely.

The electronic band structures of \textit{\textbf{ds}}-CsPb$_{2}$Br$_{5}$ with V$_{Cs}$, V$_{Br_{i}}$, V$_{Br_{s}}$ and V$_{Pb}$ are presented in Figs. \ref{fig-4} (a)-(d), respectively. It appears that electronic properties of single-layer \textit{\textbf{ds}}-CsPb$_{2}$Br$_{5}$ do not change significantly with the removal of Cs atom. Band decomposed charge densities at the valence band and conduction band edges of V$_{Cs}$ show similar behavior with that of pristine \textit{\textbf{ds}}-CsPb$_{2}$Br$_{5}$. Figs. \ref{fig-4} (b) and (c) display band decomposed charge densities and 
electronic band diagrams of V$_{Br_{i}}$ and V$_{Br_{s}}$. It is seen that only the formation of 
Br vacancies leads to presence of in-gap states. Such dispersionless electronic states  
can be attributed to the unpaired electrons of neighboring Pb and Br atoms. For V$_{Br_{i}}$ and V$_{Br_{s}}$ vacancies, 
two spin-polarized in-gap states are symmetrically placed around the Fermi level 
with band gaps of 0.12 and 0.25 eV, respectively, which then describe a singly 
occupied in-gap state with a magnetic moment of 1 $\mu_{B}$. Fig. 
\ref{fig-4} (d) shows that when formation of Pb vacancy is created, some states of 
\textit{\textbf{ds}}-CsPb$_{2}$Br$_{5}$ discharge due to the lack of donor electrons of Pb atom. As a 
result, formation of Pb vacancy leads to p-type conductivity in single-layer 
\textit{\textbf{ds}}-CsPb$_{2}$Br$_{5}$.

\section{Conclusions}\label{sec6}

In conclusion, using first-principle calculations, we investigated the structural, electronic and vibrational properties of CsPb$_{2}$Br$_{5}$ crystal, and how these properties are affected by dimensional crossover. Bulk  CsPb$_{2}$Br$_{5}$ is an indirect bandgap semiconductor with a bandgap of 2.41 (LDA+SOC) eV. It was calculated that while the valence and conduction band edges of bulk  CsPb$_{2}$Br$_{5}$ crystal are mainly composed of Pb and Br atoms, Cs atoms do not play role in electronic properties.

Then, we predicted that there are two dynamically stable phases of single layer CsPb$_{2}$Br$_{5}$. These two structures were found to be stable in both total energy optimization and phonon calculations. Single-layer structures of CsPb$_{2}$Br$_{5}$ display indirect semiconducting character with a band gap of $\sim$ 2.54 eV within LDA+SOC. Moreover, we showed that stability of single layer CsPb$_{2}$Br$_{5}$ structures strongly depend on the concentration of Cs atoms. As supported by charging-dependent phonon dispersion calculations, Cs atoms provide stability by 0.8 $e$ per atom charge transfer from Cs atoms to the Pb$_{2}$Br$_{5}$ skeleton.

In addition, formation characteristics, electronic structure, and magnetic ground state of four different vacancy types in single-layer CsPb$_{2}$Br$_{5}$ were investigated. It was seen that the formation of Br vacancy is the most likely one and it leads to emergence of localized in-gap states. Moreover, single-layer \textit{\textbf{ds}}-CsPb$_{2}$Br$_{5}$ has a ferromagnetic ground state upon the formation of Br vacancies. On the other hand, p-type doping occurs in semiconducting single-layer CsPb$_{2}$Br$_{5}$ when Pb and Cs vacancies are formed.

\section{acknowledgments}
Computational resources were provided by TUBITAK ULAKBIM, High Performance and Grid Computing Center (TR-Grid e-Infrastructure).  HS acknowledges financial support from the TUBITAK under the project number 116C073. HS acknowledges support from Bilim Akademisi-The Science Academy, Turkey under the BAGEP program.

\end{document}